\documentclass[12pt]{article}
\usepackage[cp1251]{inputenc}
\usepackage[dvips]{graphics}
\usepackage{epsfig}

\usepackage[centerlast]{caption}
\begin{document}
\large
\title{The interparticle interaction and noncommutativity
of conjugate operators in quantum mechanics. $H$-like atoms.}
        \author{A.I.Steshenko \\ \\
\centerline{\it Bogolyubov Institute for Theoretical Physics of
the NAS of Ukraine,}  \\ \it Metrolohichna Str., 14b, Kyiv-143,
03143, Ukraine }
\date{}
\maketitle

\begin{quotation}
   A quantum mechanical model for the systems consisting of
interacting bodies is considered. The model takes into account the
noncommutativity of the space and impulse operators and the
correlation equations for the indeterminacy of these quantities.
The noncommutativity of the operators is here a result of the
action of the interparticle forces and represents a natural
generalization of the conventional commutation relation for the
space and impulse operators for a single particle. The efficiency
of the model is demonstrated by specific calculations concerning
several well-known atomic systems.

\end{quotation}

\section{Introduction.}

\quad \,Earlier, in Refs. ~\cite{1} - ~\cite{2}, there has been
put forward the idea that the coordinate and impulse operators for
different particles may be not commutative, based on the following
arguments: "Abandoning the implicit assumption that the
interactions can propagate with finite velocity results in the
noncommutativity of the coordinate and impulse operators for
different particles". In this paper, we develop this idea, namely,
we give a little different physical substantiation for the fact of
noncommutativity of the above operators (as compared to Refs.
~\cite{1} - ~\cite{2}), and introduce the correlation equations
(CE) for the indeterminacy of coordinates and impulses of
different particles. This latter circumstance (i.e., introducing
CE) has, actually, changed the meaning of the model from
exclusively "theoretical and philosophical" to the " theoretical
and applied" one, and, therefore, made it possible to perform the
high precision specific calculations ~\cite{1} - ~\cite{2}.

Below, within the framework of the NOCE model (the NOCE model
means the noncommutativity of the operators and the correlation
equations), we are going to write and examine in detail the
equations for the ground and some excited states of Hydrogen-like
($H$-like) atoms.

Before we proceed to formulating the NOCE model, let us mention
that, in quantum mechanics, the many-particle Schrodinger equation
(SE) for non-interacting particles and SE for a system of
interacting particles differ only in the presence or absence of
the terms containing the potential $V_{ij}$. In actual fact,
however, the case with the interaction is fundamentally different
from the one with $V_{ij}\equiv0$. As follows from the analysis,
the many-particle SE alone is insufficient for a more accurate
description of quantum systems with $V_{ij}\equiv0$. Let us give a
try to briefly formulate the basic points of this analysis
starting at the moment of the "emergence"\, of SE.

As is well known, in order to write SE, one needs to apply the
formal transformation
\begin{equation}
\label{1} E \rightarrow i \hbar \frac{\partial}{\partial t},
\qquad p_j \rightarrow-i\hbar\frac{\partial}{\partial q_j},
\end{equation}
to the classical system
\begin{equation}
\label{2} E=H(p_j ,q_j)
\end{equation}
under consideration. Here $p_j, q_j $ is a pair of the canonically
conjugate coordinates (the impulse and space coordinate) of the
$j-$th particle. To avoid ambiguity, it is understood (see, for
instance, ~\cite{3}) that the transformation (1) has to be applied
only in the case that the independent coordinates $q_j$ are the
Cartesian ones. It is clear that obtaining the operator equation
for the wave function $\Psi(1,2,...,A)$ in such a way may not be
considered as a rigorous deduction of the equation of motion
(meaning the SE); the latter is, as mentioned in ~\cite{4}, "the
generalization of the experimental facts".

One of the important results of such a quantum mechanical
"generalization of the experimental facts" is the commutation
relation for the operators of the generalized coordinate
$\widehat{q}$ and its conjugate impulse $\widehat{p}$
\begin{equation}
\label{3} [\widehat{q},\widehat{p}]=i\hbar\,.
\end{equation}
This basic relation is valid for arbitrary quantum objects of
microcosm. What physics underlies Eq.(3)? This question, which
attracted attention of the founders of the quantum theory (see,
for instance, ~\cite{5} -~\cite{7}) remains of interest nowadays
as well (in this connection we can cite the original and to some
extent unexpected results of A.D.Sukhanov ~\cite{8} -~\cite{9}).
The commonly accepted interpretation of the relation (3) amounts
to the statement that the physical quantities $q, p$ associated
with the operators $\widehat{q}$ and $\widehat{p}$ can be found
simultaneously only with the accuracy
\begin{equation}
\label{4} \triangle q \triangle p \geq \frac{\hbar}{2}.
\end{equation}
In other words, the inaccuracies in the measurements of the
impulse and coordinate of a particle turn out to be the correlated
ones.

Up to now, we considered the conjugate coordinates of the same
(single) particle, where the situation appears rather clear.
Consider now a set of $A$ particles interacting due to some
potential $V$. A question arises, is there a correlation in the
inaccuracies in the simultaneous measurements of the impulses and
coordinates, provided that these latter correspond to
\underline{different} particles? As is known, the conventional
theory gives here a negative answer, which is, in our opinion, not
quite correct. Really, let us admit, for instance, that the
interaction between the particles $1$ and $2$ is strong to the
extent that these latter can be observed in experiments as a
single massive particle. In this case the presence of the
correlation between the inaccuracies in the measurements of the
coordinate of the $1$-st particle and the impulse of the $2$-nd
particle is beyond any doubt. For a weak interaction, these
correlations may be almost invisible in experiments, however, in
principle, they must exist.

\section{ The formulation of the model.}  \label{sect2}

After this introduction, let us formulate the NOCE model. First,
for the sake of simplicity, we consider the case of two
interacting quantum particles with the masses $m_1$ and $m_2$,
respectively, measured before they form  the bound system. Let us
begin with the commutation relations for the conjugate coordinates
$\mathbf{r_1}, \mathbf{r_2}$, and $\mathbf{p_1}, \mathbf{p_2}$. In
the NOCE model they read
\begin{equation}
\label{5} [x_k,\hat{p}_l^x]=[y_k,\hat{p}_l^y]=[z_k,\hat{p}_l^z]=
  i \hbar \beta_{kl}, \qquad  k,l=1,2;
  \end{equation}
where $(x_k,y_k,z_k)$ - are the Cartesian coordinates of the
$k$-th particle, and $ \hat{\mathbf{p}}_l
(\hat{p}_l^x,\hat{p}_l^y,\hat{p}_l^z)$ is the associated impulse
operator of the $l$-th particle, more exactly,
\begin{equation}
\label{6} \mathbf{\hat{p}_1}=-i\hbar
\beta_{11}\cdot\mathbf{\bigtriangledown}_1,\qquad\mathbf{\hat{p}_2}=-i\hbar
\beta_{22}\cdot\mathbf{\bigtriangledown}_2\,.
\end{equation}

It should be noted that the Planck constant $\hbar$ and the
quantities $\beta_{kl}$ play here the role of the commutation
parameters of the theory. The numerical values for $\beta_{kl}$
depend, in contrast to $\hbar$, on the nature of the specific
particles under consideration. The range for the quantities
$\beta_{kl}$ represents, according to the physical sense, the
interval $[0 < \beta_{kl} \leq 1]$; also, it is clear that in the
case $k \neq l$ the strong inequality $\beta_{kl} \ll \beta_{ll}$
holds. The commutation relation for the operator of the total
impulse of the system
$\mathbf{\hat{P}}=\mathbf{\hat{p}_1}+\mathbf{\hat{p}_2}\,$ and the
space coordinates of the particles can be written in the form
~\cite{1}:
\begin{equation}
\label{7}  [x_k,\hat{P}^x]=[y_k,\hat{P}^y]=[z_k,\hat{P}^z]=i\hbar
,\qquad k=1,2.
\end{equation}
Examine now the Hamilton function $H$. For two classical particles
it reads
\begin{equation}
\label{8} H=\frac{p_1^2}{2m_1}+\frac{p_2^2}{2m_2}+V_{12}\, ,
\end{equation}
i.e., taking into account (6), we obtain the following SE
\begin{equation}
\label{9}
\left[-\frac{\hbar^2}{2(m_1/\beta_{11}^2)}\cdot\mathbf{\bigtriangledown}_1^2
-\frac{\hbar^2}{2(m_2/\beta_{22}^2)}\cdot\mathbf{\bigtriangledown}_2^2+
V_{12}\right ]\Psi(1,2)=i\hbar\frac{\partial\Psi}{\partial t}\,.
\end{equation}
As can be seen, Eq(9) differs from the conventional SE only by the
modifications in the particle masses
\begin{equation}
\label{10} m_1^{'} \equiv \frac{m_1}{\beta_{11}^2},\qquad
m_2^{'}\equiv\frac{m_2}{\beta_{22}^2}\,.
\end{equation}

The next problem should be, evidently, the determination of these
modified masses $m^{'}_1$ and $m^{'}_2$. For this purpose, we
consider the process of the measurement of the coordinate of the
1-st particle $x_1$, which is carried out with the maximum
precision (this is a common treatment for evaluating the quantum
mechanical averages; see, for instance ~\cite{10}). The maximum
precision in the measurement of the coordinate of the particle is,
apparently, limited by its Compton wave length, i.e.,
\begin{equation}
\label{11} \Delta x_1=\frac{\hbar}{m_1 c}\,.
\end{equation}
In the course of the measurement of $x_1$, the second particle
acquires, due to the interaction between the 1-st and the 2-nd
particles, the impulse
\begin{equation}
\label{12} \Delta p_2^x=\langle |F|\rangle\cdot\Delta t \,,
\end{equation}
where $\langle \, |F| \,\rangle \equiv \langle
\Psi(1,2)|F_{12}|\Psi(1,2)\rangle$ is the matrix element of the
force calculated with the wave functions $\Psi(1,2)$, and the
quantity $\Delta t$ is the interaction time. For the latter, one
can take the so-called "passing time"\, for the 1-st particle
\begin{equation}
\label{13} \Delta t=\frac{\Delta x_1}{c} \,.
\end{equation}
Taking into account the relations (11) and (13), we rewrite (12)
in the form
\begin{equation}
\label{14} \Delta p_2^x=\frac{\hbar}{m_1c^2} \, \langle |F|\rangle
\,.
\end{equation}
For further consideration, it is advisable to examine, along with
the Heisenberg indeterminacy relations
\begin{equation}
\label{15} \Delta x_k\Delta p^x_l=\Delta y_k\Delta p^y_l=\Delta
z_k\Delta p^z_l\geq \frac{\hbar}{2} \beta_{kl}\, ; \qquad
k,l=1,2\, ,
\end{equation}
the corresponding correlation equations
\begin{equation}
\label{16} \gamma_{kl}\Delta x_k\Delta p^x_l=\gamma_{kl}\Delta
y_k\Delta p^y_l=\gamma_{kl}\Delta z_k\Delta p^z_l= \frac{\hbar}{2}
\beta_{kl}\, , \qquad k,l=1,2\,.
\end{equation}
Then, the impulse of the 1-st particle obtained by the measurement
of the coordinate $x_1$ equals, according to (16) and (11),
\begin{equation}
\label{17} \Delta p_1^x=\frac{\hbar \beta_{11}}{2\,\gamma_{11}
\Delta x_1}=\frac{m_1 c}{2} \cdot \frac{\beta_{11}}{\gamma_{11}}
\,.
\end{equation}
The magnitude of the correlation factors $\gamma_{kl}$ may not be
equal to zero, as follows from the definition (16); for this
reason the division by $\gamma_{kl}$ in (17) is quite acceptable.
As for the range of possible values for $\gamma_{kl}$, it is
evidently identical with the range for commutation parameters
$\beta_{kl}$.

Equating the expressions for the ratio $(\Delta p_2^x/\Delta
p_1^x)$ obtained, on the one hand, from Eqs.(14), (17), and, on
the other hand, from the correlation equations (16), yields
\begin{equation}
\label{18} \frac{\beta_{12}}{\gamma_{12}}=\frac{2 \hbar
c}{\varepsilon_1^2} \, \langle |F|\rangle , \qquad \varepsilon_1
\equiv m_1c^2\,.
\end{equation}
The second equality needed has to be found similarly to Eq.(18)
provided  that the particles 1 and 2 exchange their roles. Thus,
we obtain
\begin{equation}
\label{19} \frac{\beta_{21}}{\gamma_{21}}=\frac{2 \hbar
c}{\varepsilon_2^2} \, \langle |F|\rangle , \qquad \varepsilon_2
\equiv m_2c^2\,.
\end{equation}
One more pair of equations, which establishes the connection
between the diagonal $(k=l)$ and nondiagonal $( k \neq l)$
quantities $\beta_{kl}$, can be found from the commutation
relations (5) and (7),
\begin{equation}
\label{20} \left\{ \begin{array}[c]{lcr} \beta_{11}+ \beta_{12}
=1\, ;\\ \beta_{21}+\beta_{22}=1\,.
\end{array} \right.
\end{equation}
Finally, the equations (18),(19), and (20) yield the sought-for
commutation parameters $\beta_{11}$ and $\beta_{22}$ as a function
of the matrix element (ME) of the force $\langle |F|\rangle$ and
the correlation factors $\gamma_{12},\, \gamma_{21}$
\begin{equation}
\label{21} \left\{ \begin{array}[c]{lcr} \beta_{11}=1-
\frac{2\hbar c}{\varepsilon_1^2}\cdot\gamma_{12}\, \langle
|F|\rangle, \\\\ \beta_{22}=1- \frac{2\hbar
c}{\varepsilon_2^2}\cdot\gamma_{21}\,\langle |F|\rangle \,.
\end{array} \right.
\end{equation}
As a result, the problem of two interacting particles is reduced
to solving SE (9) with particle masses $m_1^{'}$ and $m_2^{'}$
being dependent, in their turn, on the unknown solution of SE,
i.e., the wave function $\Psi(1,2)$.

Note that the NOCE model under consideration can be applied to the
stationary states $\Psi(1,2)=exp[i\frac{E}{\hbar}\, t]\cdot
\psi(\mathbf{r}_1,\mathbf{r}_2)$ only. In general case including
the nonstationary states, this model needs essential modification.
The basis of the theory should, as mentioned in Refs.~\cite{8}
-~\cite{9}, probably, constitute the Schr$\ddot{o}$dinger
indeterminacy relations
\begin{equation}
\label{22} (\Delta q)^2(\Delta p)^2\geq |\tilde{R}_{qp}|^2\, ,
\end{equation}
instead of the Heisenberg indeterminacy relations (4). Here the
generalized correlator $|\tilde{R}_{qp}|=\sqrt{\sigma
_{qp}^2+c_{qp}^2}$ represents a complex number with the imaginary
part  $c_{qp}=\frac{\hbar}{2}$ (for the stationary states
$\sigma_{qp}=0$).

Thus, the bound states of two particles are described in the NOCE
model by the following equations:
\begin{equation}
\label{23}  \left\{ \begin{array}[c]{lcr}
\left[-\frac{\hbar^2}{2m_1^{'}}\cdot\mathbf{\bigtriangledown}_1^2
-\frac{\hbar^2}{2m_2^{'}}\cdot\mathbf{\bigtriangledown}_2^2+
V_{12}\right ]\psi(1,2)=E\cdot\psi(1,2);\quad
m_1^{'}=\frac{m_1}{\beta_{11}^2},\,
m_2^{'}=\frac{m_2}{\beta_{22}^2 }\, ; \\\\
 \beta_{11}=1- \frac{2\hbar
c}{\varepsilon_1^2}\cdot\gamma_{12}\langle |F|\rangle\,, \quad
\beta_{22}=1- \frac{2\hbar
c}{\varepsilon_2^2}\cdot\gamma_{21}\langle |F|\rangle\,.
\end{array} \right.
\end{equation}
Specific solutions to these equations can be found by the method
of successive iterations, where at the 1-st step the conventional
SE with the masses $m_1$ and $m_2$, which the particles have in
the absence of the interaction ($\beta_{11}=\beta_{22}\equiv1$),
has to be solved. After that, on finding the wave function $\psi$,
one can calculate ME of the force $\langle |F|\rangle$ and the
first values of the commutation parameters $\beta_{11}$ and
$\beta_{22}$ distinct from unity. At the 2-nd step, SE is solved
with the modified particle masses $m_1^{'}=m_1/\beta_{11}^2$ and
$m_2^{'}=m_2/\beta_{22}^2$. On finding the new $\psi$, we
calculate the quantity  $\langle |F|\rangle$ and compare it with
the one obtained at the 1st step. Then, we proceed with the
iterations until the values of ME of the force $\langle
|F|\rangle$ obtained at subsequent steps will be virtually
indistinguishable. It is clear that, before starting the above
iteration process, we should specify the numerical values for the
correlation factors $\gamma_{12}$ and $\gamma_{21}$ entering
Eqs.(23). To calculate them, one can employ specific parameters of
a given system based on reliable experimental data. The way to
practically implement this will be described in detail
hereinafter.

\section{Hydrogen-like atoms } \label{sect3}

\qquad Let us first apply the NOCE model for two interacting
bodies under consideration  to Hydrogen atom and some similar
atoms ($H$-like atoms). For stationary states, the system of
equations (23) has the form
\begin{equation}
\label{24}  \left\{ \begin{array}[c]{lcr}
[-\frac{\hbar^2}{2m_1^{'}}\cdot\mathbf{\bigtriangledown}_1^2
-\frac{\hbar^2}{2m_2^{'}}\cdot\mathbf{\bigtriangledown}_2^2+\frac{Ze^2}
{|\mathbf{r}_1-\mathbf{r}_2|}]\psi(1,2)=E\cdot\psi(1,2)\,; \\\\
m_1^{'}=m_e/(1- \frac{2\hbar c}{\varepsilon_1^2}\,\langle
|F|\rangle\cdot\gamma_{12})^2,\qquad m_2^{'}=m_H/(1- \frac{2\hbar
c}{\varepsilon_2^2}\, \langle |F|\rangle\cdot\gamma_{21})^2 \,,
\end{array} \right.
\end{equation}
where $m_e, m_H$ are the electron and $Z$-nucleus masses,
respectively; $\langle |F|\rangle\equiv \langle
\psi(1,2)|\frac{Z\,e^2}
{|\mathbf{r}_1-\mathbf{r}_2|}|\psi(1,2)\rangle$  is ME of the
force. The first equation in (24) for fixed $m_1^{'}$ and
$m_2^{'}$ is described in detail in a number of reference books on
quantum mechanics. Introducing the variables
\begin{equation}
\label{25}   \mathbf{r}=\mathbf{r}_1-\mathbf{r}_2\, ,\qquad
\mathbf{R}_{c.m.}=\frac{m_1^{'}\mathbf{r_1}+m_2^{'}\mathbf{r_2}}{m_1^{'}+m_2^{'}}
\end{equation}
results in the factorization of the wave function
$\psi(1,2)=\Phi(\mathbf{R}_{c.m.})\cdot\psi(\mathbf{r})$. Of
physical interest therewith is the function
\begin{equation}
\label{26}
\psi(\mathbf{r})\equiv\psi_{nlm}(r,\theta,\varphi)=R_{nl}(r)\cdot
Y_{lm}(\theta,\varphi)\, ,
\end{equation}
only, with $Y_{lm}(\theta,\varphi)$ being the spherical functions
~\cite{11}. The function $R_{nl}(r)$ has to be found from the
so-called radial SE
\begin{equation}
\label{27}
\left[\frac{d^2}{dr^2}+\frac2r\cdot\frac{d}{dr}+\frac{2\mu^{'}}{\hbar^2}
\left(E+\frac{Ze^2}{r}\right)-\frac{l(l+1)}{r^2}\right]R_{nl}(r)=0\,.
\end{equation}
The latter yields for the bound states the solution (see, for
instance ~\cite{12})
\begin{equation}
\label{28}  \begin{array}{ll}
 R_{nl}(r)=\left(\frac{Z}{na_o}\right)^{3/2}
\frac{2\cdot n!}{\sqrt{n(n-l-1)!(n+l)!}}\left(\frac{2Zr}{na_o}\right)^l
\exp{\left(\frac{-Zr}{na_o}\right)}L_{n-l-1}^{2l+1}
\left(\frac{2Zr}{na_o}\right),\\ n=1,2,...;\, l=0,1,2,...,n-1\, ,
\end{array}
\end{equation}
where $ L_n^\alpha$ are the generalized Laguerre polynomials
~\cite{13}, $\mu^{'}$ is the modified reduced mass, $a_o$ is the
radius of the 1-st Bohr orbital, which are, respectively, given by
\begin{equation}
\label{29}  \begin{array}{ll}
 L_n^\alpha (x)= \frac{1}{n!}\, e^x x^{-\alpha}\, \frac{d^n}{dx^n}
 (e^{-x}\,x^{n+\alpha})=
 \sum_{k=0}^{n}\frac{\Gamma (n+\alpha+1)}
 {\Gamma(k+\alpha+1)}\cdot\frac{(-x)^k}{k! (n-k)!};\\\\ \mu^{'}=
 \frac{m_1^{'}m_2^{'}}{m_1^{'}+m_2^{'}};\qquad
a_o=\frac{\hbar^2}{\mu^{'}e^2}.
\end{array}
\end{equation}
The explicit form of the known equations given in the relations
(27)-(29) is needed to find some algebraic equation. Its solution,
in this specific case, can make the above iteration procedure
unnecessary. For this purpose, we write ME of the force $\langle
|F|\rangle$ on the functions $\psi_{nlm}$
\begin{equation}
\label{30} \langle |F|\rangle=\langle
\psi_{nlm}|\frac{Ze^2}{r^2}|\psi_{nlm}\rangle=\frac{Z\,e^2}{n^3\,
(l+\frac12)}\,\left(\frac{Z}{a_o}\right)^2\equiv f_{nl}\,.
\end{equation}
For the ground state
\begin{equation}
\label{31}
\psi_{100}=\frac{1}{\sqrt{\pi}}\left(\frac{Z}{a_o}\right)^{3/2}
\exp\left(-\frac{Zr}{a_o}\right)
\end{equation}
ME of the force is $f_{10}=2\,Z\,e^2(Z/a_o)^2$.

As is seen from the above relations, the quantity $f_{nl}$ depends
on the Bohr radius $a_o$, as well as the wave function
$\psi_{nlm}$. The Bohr radius, in turn, is the function of
$f_{nl}$. Really, the relations (29) and (24) yield
\begin{equation}
\label{32}
a_o=\frac{\hbar^2}{\mu^{'}e^2}=\frac{\hbar^2}{e^2}\left[\frac{1}{m_1}
\left(1-\gamma_{12}\frac{2\hbar c}{\varepsilon_1^2}\cdot
f_{nl}\right)^2 + \frac{1}{m_2} \left(1-\gamma_{21}\frac{2\hbar
c}{\varepsilon_2^2}\cdot f_{nl}\right)^2 \right] \, .
\end{equation}
By substituting this expression into the formula (30), we obtain
the sought-for equation for $f_{nl}$
\begin{equation}
\label{33}   \frac{(2l+1)\,n^3}{4(\alpha\,Z)^3}\cdot\frac{2\hbar
c}{\varepsilon_1^2}\, f_{nl}= \frac{1}{\left[
\left(1-2\,\gamma_{12}\,\frac{\hbar c}{\varepsilon_1^2}\,
f_{nl}\right)^2+\xi \,\left(1-2\,\gamma_{21}\,\xi^2\, \frac{\hbar
c}{\varepsilon_1^2} \, f_{nl}\right)^2\right]^2} \, ,
\end{equation}
where $\alpha\equiv e^2/\hbar c$ is the fine structure constant
and the quantity $\xi\equiv \varepsilon_1/\varepsilon_2$. The
equation (33) enables us to determine ME of the force $f_{nl}$
without any iteration procedures and to find after that the
magnitudes of the "inertial" particle masses $m_1^{'}$ and
$m_2^{'}$, which have to be substituted into SE.

The energy spectrum of the bound states of the $H$-like atoms is
given by the equation
\begin{equation}
\label{34} E_n=-\frac12\left(\frac{\alpha \,
Z}{n}\right)^2\frac{\varepsilon_1}{\left(\beta_{11}^2+\xi\cdot\beta_{22}^2\right)}\,
.
\end{equation}
To perform specific calculations, it is necessary to specify the
numerical values for the correlation factors $\gamma_{12}$ and
$\gamma_{21}$ as well as the values for all physical constants
entering Eqs.(24)-(29). In the limiting case
$\beta_{11}=\beta_{22}=1$ we obtain, apparently, the relations of
the conventional quantum mechanics. In this case the energy of the
ground state of Hydrogen atom $(Z=1)$
\begin{equation}
\label{35}
E_{n=1}=-\alpha^2\frac{\varepsilon_1}{2}\cdot\frac{1}{1+\varepsilon_1/\varepsilon_2}\,
,
\end{equation}
for the values $\varepsilon_1=0.510998902\, MeV$,
\,$\varepsilon_2=938.271998\, Mev$, and $\,
\alpha^{-1}=137.03599976 $ turns out to be
$E_1^{'}$=-13.598\,285\,8517 eV. In the other limiting case
$\gamma_{12}=\gamma_{21} =1$ the equation (33) takes the form
\begin{equation}
\label{36}
x\left[(1-2\,x)^2+\xi(1-2\xi^2\,x)^2\right]^2=2\,\alpha^3\, .
\end{equation}
Here the dimensionless quantity
\begin{equation}
\label{37} x\equiv \frac{\hbar c}{\varepsilon_1^2}\,f_{10}
\end{equation}
is introduced. The equation (36) has one positive root
$x_o=7,763\,467\,765\,809\cdot10^{-7}$, which gives for ME of the
force $f_{10}=\varepsilon_1^2\,x_o/\hbar c \approx
1,027\,328\,325\cdot 10^{-9} $ MeV/fm. For this value $f_{10}$ we
obtain  $\beta_{12}=1-\beta_{11}=\frac{2\,\hbar
c}{\varepsilon_1^2}\cdot f_{10} = 0,155\,269\,355
\cdot10^{-5},\,\beta_{21}=1-\beta_{22}=\frac{2\,\hbar
c}{\varepsilon_2^2}\cdot
f_{10}=0,460\,540\,866\cdot10^{-12},\,\triangle m_e \, c^2\equiv
m_e^{'}\,c^2-m_e\,c^2= 0,158\,685\,310\cdot10^{-5}$ MeV,
$\triangle m_H\,c^2 \equiv m_H^{'}\,c^2-m_H\,c^2=
0,864\,225\,197\cdot10^{-9}$, and the energy
$E_{n=1}^{''}=-13,598\,328\,057$ eV. Taking for the experimental
energy the value
\begin{equation}
\label{38}
E_1^{exp}=-R_{\infty} hc\cdot \frac{1}{1+
\frac{\varepsilon_1}{\varepsilon_2}}\, ,
\end{equation}
with  $R_{\infty} hc=13.605\,691\,72(52)\,$ eV, in accordance with
Ref.~\cite{14}, one can see that the value
$E_1^{exp}=-13.598\,285\,862\,$ eV comes to the interval formed by
the above mentioned limiting estimates $[E_1^{'},\,E_1^{''}]$ in
the vicinity of the point $E_1^{'}$.

To calculate the correlation factors $\gamma_{12}$ and
$\gamma_{21}$, we use the value for the energy of the ground state
of the Hydrogen atom (the ionization energy of the atom)
\begin{equation}
\label{39} E_1^{exp.}=\frac{R_{\infty} hc}{1+
\xi}=\frac{\varepsilon_1\,\alpha^2}{2}\frac{1}{(1-2\frac{\hbar c}
{\varepsilon_1^2}f_{10}\cdot\gamma_{12})^2
+\xi(1-2\xi^2\frac{\hbar
c}{\varepsilon_1^2}f_{10}\cdot\gamma_{21})^2}\, ,
\end{equation}
by taking as ME of the force $f_{10}$ its approximate value
\begin{equation}
\label{40} f_{10}\rightarrow
\tilde{f}_{10}=2\,\frac{\alpha^3\varepsilon_1^2}{\hbar c}\cdot
\frac{1}{(1+\xi)^2}\,.
\end{equation}
Then, the expression for $\gamma_{12}$ reads
\begin{equation}
\label{41}
\gamma_{12}=\frac{(1+\xi)^2}{4\,\alpha^3}\left[1-\sqrt{\frac{\varepsilon_1\,
\alpha^2\,(1+\xi)}{2\cdot R_{\infty}hc}-\xi\left (1-4\alpha^3\,
\frac{\xi^2}{(1+\xi)^2}\cdot\gamma_{21}\right)^2}\,\right] \, .
\end{equation}

Note that the theory of the Hydrogen atom under consideration
turns out to de indifferent in relation to the second correlation
factor $\gamma_{21}$ due to the smallness of the coefficient
$4\alpha^3\cdot\xi^2/(1+\xi)^2\simeq 4.605\cdot 10^{-13}$. Running
the values of $\gamma_{21}$ from $\gamma_{21}=0.0001$ to
$\gamma_{21}=1.0$ yields the difference for the quantity
$\gamma_{12}$ in the 7-th digit only. At the same time, the
quantity $\gamma_{12}$ is extremely sensitive to the value of the
Rydberg constant $R_{\infty}hc$. In particular, for the
experimental value $R_{\infty}hc$=13.605\,691\,72 eV  ~\cite{14}
we obtain $\gamma_{12}\simeq 2.44463\cdot10^{-4}$. However, if we
restrict ourselves to the approximate (6 digits only) value
$R_{\infty}hc$=13.605\,7 eV, then $\gamma_{12}\simeq 0.19632$, and
for the "five-digit" approximation for the Rydberg constant
$R_{\infty}hc$=13.606 eV we obtain the value $\gamma_{12}\simeq
7.3$, which makes no sense within the framework of the present
model $(\,[0<\gamma_{12}\leq 1]\,)$. In other words, in the case
that the Rydberg constant turned out to be only 0.0003 eV greater,
our theory had to be discarded as the one contrary to the fact.

We now turn our attention to the algebraic equation (33). We
substitute into it the estimate
$\gamma_{12}=\gamma_{21}\equiv\gamma\approx 2.44463\cdot10^{-4}$
found for Hydrogen atom instead of correlation factors and then
rewrite  this equation in the form of the polynomial in the powers
of the unknown $x\equiv \frac{\hbar c\cdot
f_{10}}{\varepsilon_1^2}$, i.e.,
\begin{equation}
\label{42} a_1\cdot x-a_2\cdot x^2+a_3\cdot x^3-a_4\cdot
x^4+a_5\cdot x^5-2\,\alpha^3=0\, ,
\end{equation}
with the coefficients $a_i$ given by
\begin{equation}
\label{43} \begin{array}{ll} x\equiv\frac{\hbar
c}{\varepsilon_1^2}\cdot f_{10}\,, \qquad
2\,\alpha^3=0.777\,187\,805\,862\cdot 10^{-6}\, ,\\\\
a_1=(1+\xi)^2\approx 1.001\,089\,530\,653\, ,\\\\
a_2=8\gamma(1+\xi)(1+\xi^3)\approx1.956\,770\,707\,906\cdot
10^{-3}\, ,\\\\
a_3=16\gamma^2[(1+\xi^3)^2+\frac12(1+\xi)(1+\xi^5)]\approx
1.434\,554\,523\,990 \cdot10^{-6}\, ,\\\\
a_4=32\gamma^3(1+\xi^3)(1+\xi^5)\approx
4.675\,095\,140\,637\cdot10^{-10}\, ,\\\\
a_5=16\gamma^4(1+\xi^5)^2\approx 5.714\,443\,582\,322\cdot
10^{-14}
 \, .
\end{array}
\end{equation}
The intersection points with the x-axis for the function
$\varphi(x)=a_1\,x-a_2\,x^2+a_3\,x^3-a_4\,x^4+a_5\,x^5-2\,
\alpha^3$ represent, evidently, the roots $x_i\equiv f_{10}^{(i)}$
of the equation under consideration (42).In Fig.1, the general
plot of the function $\varphi(x)$ is displayed, as well as the
behavior of the curve $\varphi(x)$ in the vicinity of the points
$x_1=8\cdot10^{-7}$ and $x_2=2050.0$ on a magnified scale. As is
seen from this figure, there is definitely one
root$x_1=7.763\,419\,586\,802\cdot 10^{-7}$.The associated value
for ME of the force $f_{10}^{(1)}= \frac{\varepsilon_1^2}{\hbar c}
\cdot x_1= 1.027\,321\,949\,956\cdot10^{-9}$ MeV/fm turned out to
be very close to the approximate estimate (40)
$\tilde{f}_o=1.027\,321\,947\,815\cdot10^{-9}$ MeV/fm. As for
other possible positive roots of the polynomial $\varphi(x)$, one
should take into account that their number, according to the
Decarte's rule, may differ from the number of sign changes in the
polynomial $\varphi(x)$ by an even number only, hence, in the
present case the number of positive roots may be 1, 3, or 5 only.
However, for the above given fundamental constants there exists
only one root. The deviation of the electron mass for ME of the
force $f_{10}=f_{10}^{(1)}$ is rather small, it equals $\triangle
m_e\,c^2\equiv m_e^{'}\,c^2-m_e\,c^2 \approx
0,387\,924\cdot10^{-9}$ MeV.
\begin{figure}[t]
\centering{\includegraphics[width=11cm]{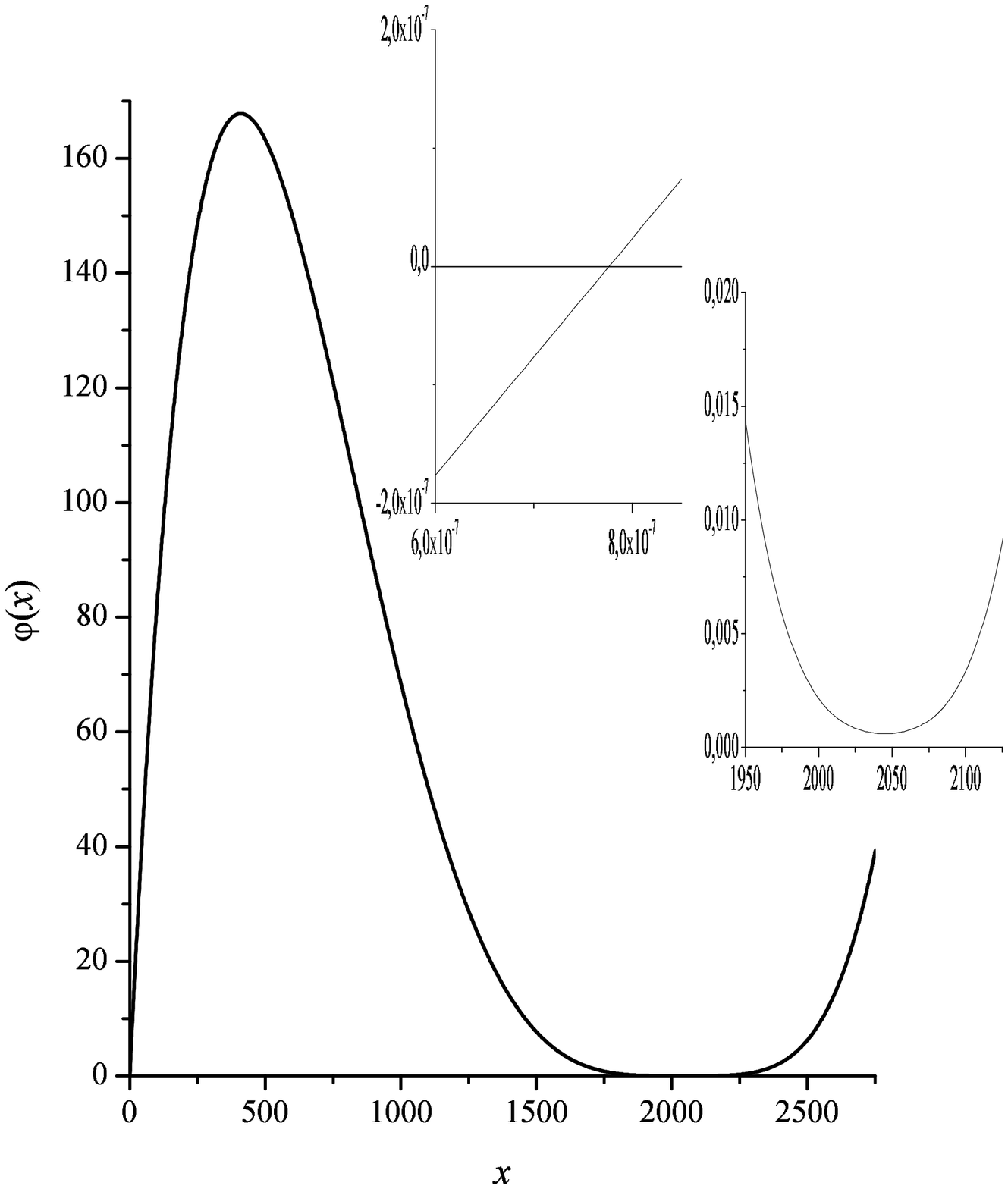}} \caption{\it{ The
plot of the function $\varphi(x)$ (see the text).}}
 \label{fig.1}
\end{figure}
Let us apply the theory of $H$-like atoms under consideration to
calculations of energy levels. Experimental data on the energy
spectra of $H$-atoms have been obtained recently ~\cite{15} -
~\cite{20} for a number of elements,  $^{12}C,\, ^{24}Mg,\,
^{40}Ar,\,^{52}Cr,\, ^{64}Zn,\, ^{84}Kr,\, ^{98}Mo,\, ^{238}U$\,
etc. The energy levels of an $H$-atom are given by the relation
\begin{equation}
\label{44} E_{nl}=-\frac{\varepsilon_1}{2}\,\frac{\alpha^2
Z^2}{n^2}\cdot\frac{1}{\left[\left(1-\frac{2\,\hbar c
}{\varepsilon_1^2}\,f_{nl}\cdot\gamma_{12}\right)^2+\xi\left(1-\frac{2\,\hbar
c}{\varepsilon_1^2}\,f_{nl}\cdot\xi^2\gamma_{21}\right)^2\right]}\,.
\end{equation}

Notice that the energy spectrum of Hydrogen-like atoms in the NOCE
model depends on $(n,l)$ due to the dependence of ME of the force
$f_{nl}$ on the two quantum numbers $n$ and $l$ (see Eqs.(30) and
(31)). In the non-relativistic quantum mechanics the energy of a
level depends, as is known, on the principal quantum number $n$
only, more exactly,
\begin{equation}
\label{45}
E_{n}^{Q.mech.}=-\frac{\varepsilon_1}{2}\,\frac{\alpha^2
Z^2}{n^2}\cdot\frac{1}{1+\xi}\,.
\end{equation}
At the same time, in the Dirac theory it depends on the total spin
$\mathbf{j}=\mathbf{l}+\mathbf{s}$ as well
\begin{equation}
\label{46} E_{nj}^{Dir.}=m_e c^2\left[1+\left(\frac{\alpha
Z}{n-j-\frac12+\sqrt{(j+\frac12)^2-\alpha^2
Z^2}}\right)^2\right]^{-1/2}\,-m_e c^2 \, .
\end{equation}
The quantity $\varepsilon_2$ entering the definition $\xi\equiv
\varepsilon_1/\varepsilon_2$ is given by the formula
$\varepsilon_2=A\cdot 1e+\bigtriangleup m_{^{12}C}$ (see, for
instance, ~\cite{21}) where $A$ is the atomic number of the
nucleus, $1e$=931.441\, MeV, and $\bigtriangleup m_{^{12}C}$ is
the excess of the mass on the $^{12}C=12$ scale given in the units
of MeV.

It is clear that, before solving the algebraic equation (33), one
needs to specify the correlation factors $\gamma_{12}$ and
$\gamma_{21}$ corresponding to the given $H$-atom. For this
purpose, we employ one or another reliable experimental value for
the energy $E_{nl}$. In doing so, ME of the force $f_{nl}$ are to
be expressed via $E_{nl}$ in accordance with the equation
\begin{equation}
\label{47}
f_{nl}=\frac{8\,n}{2l+1}\cdot\frac{E_{nl}^2}{\alpha\,Z\cdot \hbar
c }\,,
\end{equation}
that is the consequence of Eqs.(33) and (44). Then, the
correlation factor $\gamma_{12}$, which is of importance for the
theory of $H$-atoms, has the form
\begin{equation}
\label{48}\begin{array}{cl} \gamma_{12}=\frac{(2l+1)\,\alpha
Z}{16\,n}\frac{\varepsilon_1^2}{E_{nl}^2}\cdot
\\\\\cdot\left[1-\frac{\alpha\,Z}{n}\, \sqrt{\frac{\varepsilon_1}{2\,|E_{nl}|
}}\cdot \sqrt{1-\frac{\xi\,n^2}{\alpha^2 Z^2}\cdot
\frac{2\,|E_{nl}|}{\varepsilon_1}\left(1-\frac{16\,n}{2l+1}
\cdot\frac{E_{nl}^2}{\alpha Z\,\varepsilon_1^2}\,\xi^2
\gamma_{21}\right)^2}\right].
\end{array}
\end{equation}
As for the second factor $\gamma_{21}$, it is, as mentioned above,
of no concern. For the sake of definiteness, we set, in what
follows, $\gamma_{21}=\gamma_{12}$.

Now we turn to specific calculations, taking as an example the
$H$-atom $^{12}C$ which is well studied in experiments. In
Ref.~\cite{20}, there are the data on the 81-st energy level of
the $H$-atom $^{12}C$ given with accuracy $\sim 10^{-4} $ eV. In
order to determine the correlation factors, we use the
experimental value ~\cite{20} for the ground state energy of the
$H$-atom $E_{nl}=E_{10}=-489.9933$ eV. The value of the mass of
$^{12}C$ needed to determine the quantity
$\xi=\varepsilon_1/\varepsilon_2$ equals $\varepsilon_2\equiv c^2
m_{^{12}C}=11\,177.292$ MeV. Substituting these values in (48)
yields $\gamma_{12}\simeq 0.640255$. Thus, with  all the necessary
values of the physical quantities entering Eq.(33) in hand, we
proceed to calculating the energy levels of the $H$-atom $^{12}C$.
Fist, we find the roots of the algebraic (of the 5-th power with
respect to $f_{nl}$) equation (33) for the corresponding quantum
numbers $n$ and $l$. After that, by using the known value of ME of
the force $f_{nl}$, we determine the energy $E_{nl}$. In so doing,
it is convenient to use Eq.(47) instead of Eq.(44),i.e.
\begin{equation}
\label{49} E_{nl}=-\,\sqrt{\frac{2l+1}{8\,n}\cdot\alpha\,Z\cdot
\hbar c \cdot f_{nl}}\,.
\end{equation}
The result of such a calculation, along with the experimental data
available ~\cite{20}, are given in the Table 1. The energy levels
$E_{n,l}$ are calculated with respect to the ground state of the
Hydrogen-like atom $^{12}C$, more exactly,

 \mbox{}
\begin{table} []
\caption{Energy levels of the $H$-like atom, carbon-12. Energy
$(E, \delta E_i;)$ is given in eV; the dimensionless quantity
$x_{nl}$ is connected with ME of the force $f_{nl}$ via relation
$x_{nl}=\frac{\hbar c}{\varepsilon_1^2}\cdot f_{nl}$.}
\begin{center}
\begin{tabular}{|c||c|c|c|c|c|}
\hline
 $ \, n,\,l,\,j $&$E_{n,l,j}^{exp.} $ & $\delta E_1$ &$\delta E_2$
 &$\delta E_3$ &$x_{nl} $
 \\ \hline\hline
$\mathbf{1s}\, \{1,0,1/2\} $ &-$489.9933 $ &0.0000 & $ 0.2108$ &
-$0.0465 $& $0.168001731837749423\cdot 10^{-3}$
\\ \hline
$\mathbf{2p}\,\{2,1,1/2 \} $ &$367.4741$&$0.0714$& $-0.1374 $
&$0.0411$& $0.69943012924343\cdot 10^{-5}$
\\ \hline
$\mathbf{2p}\,\{2,1,3/2 \} $ & $367.5330$ & $0.0125$& $-0.1961 $&
$0.0410 $& $0.69943012924343\cdot 10^{-5}$
\\ \hline
$\mathbf{2s}\,\{2,0,1/2\} $ &$ 367.4774$  &$0.0637 $& $-0.1407 $&$
0.0378$ & $0.209844074797524\cdot10^{-4}$
\\ \hline
$\mathbf{3p}\,\{3,1,1/2\} $ & $435.5469$ &$0.0258$& $-0.1849$
 & $0.0441$ & $0.2072333324964547 \cdot 10^{-5}$
\\ \hline
$\mathbf{3p}\, \{3,1,3/2\}  $ & $435.5643 $ &$0.0084 $& $-0.2023
$& $ 0.0441$& $0.2072333324964547 \cdot 10^{-5}$
\\ \hline
$\mathbf{3s}\, \{3,0,1/2\} $ & $435.5478$  &$0.0244$& $-0.1856 $&
$0.0432$& $0.6217131956770372 \cdot 10^{-5}$
\\ \hline
$\mathbf{3d}\, \{3,2,3/2\} $ & $435.6543$ &$0.0085$& $-0.2020 $
&$0.0441$ & $0.1243394715922197\cdot 10^{-5}$
\\ \hline
$\mathbf{3d}\,\{3,2,5/2\} $ &  $435.5701$ &$0.0027$& $-0.2079 $&
$0.0441$ & $0.1243394715922197\cdot 10^{-5}$
\\ \hline
$\mathbf{4p}\,\{4,1,1/2\} $ & $459.3699$ & $0.0045$ & $-0.2062 $&
$0.0452$ & $0.87426025672203 \cdot 10^{-6}$
\\ \hline
$\mathbf{4p}\,\{4,1,3/2\} $ & $459.3773$ & $0.0119$ & $-0.1988 $&
$0.0453$ & $0.87426025672203 \cdot 10^{-6}$
\\ \hline
$\mathbf{4s}\,\{4,0,1/2\} $ & $459.3704$ & $0.0113$ & $-0.1993 $&
$0.0448$ & $0.2622804259167675\cdot 10^{-5}$
\\ \hline
$\mathbf{4d}\,\{4,2,3/2\} $ & $459.3773$ & $0.0046$ & $-0.2062 $&
$0.0452$ & $0.524555214489563\cdot 10^{-6}$
\\ \hline
$\mathbf{4d}\,\{4,2,5/2\} $ & $459.3797$ & $0.0021$ & $-0.2086 $&
$0.0452$ & $0.524555214489563\cdot 10^{-6}$
\\ \hline
$\mathbf{4f}\,\{4,3,5/2\} $ & $459.3797$ & $0.0022$ & $-0.2086 $&
$0.0452$ & $0.374682008449765\cdot 10^{-6}$
\\ \hline
$\mathbf{4f}\,\{4,3,7/2\} $ & $459.3810$ & $0.0009$ & $-0.2099 $&
$0.0452$ & $0.374682008449765\cdot 10^{-6}$
\\ \hline
$\mathbf{5p}\,\{5,1,1/2\} $ & $470.3956$ & $0.0064$ & $-0.2044 $&
$0.0457$ & $0.447620273314724\cdot 10^{-6}$
\\ \hline
$\mathbf{5p}\,\{5,1,3/2\} $ & $470.3994$ & $0.0026$ & $-0.2082 $&
$0.0457$ & $0.447620273314724\cdot 10^{-6}$
\\ \hline
$\mathbf{5s}\,\{5,0,1/2\} $ & $470.3958$ & $0.0061$ & $-0.2046 $&
$0.0455$ & $0.1342866977357545\cdot 10^{-5}$
\\ \hline
$\mathbf{5d}\,\{5,2,3/2\} $ & $470.3994$ & $0.0026$ & $-0.2082 $&
$0.0457$ & $0.268571917694502\cdot 10^{-6}$
\\ \hline
$\mathbf{5d}\,\{5,2,5/2\} $ & $470.4006$ & $0.0014$ & $-0.2094 $&
$0.0457$ & $0.268571917694502\cdot 10^{-6}$
\\ \hline
$\mathbf{5f}\,\{5,3,5/2\} $ & $470.4006$ & $0.0014$ & $-0.2094 $&
$0.0457$ & $0.191837008671438\cdot 10^{-6}$
\\ \hline
$\mathbf{5f}\,\{5,3,7/2\} $ & $470.4012$ & $0.0008$ & $-0.2100 $&
$0.0458$ & $0.191837008671438\cdot 10^{-6}$
\\ \hline
$\mathbf{5g}\,\{5,4,7/2\} $ & $470.4012$ & $0.0008$ & $-0.2100 $&
$0.0458$ & $0.149206529721493\cdot 10^{-6}$
\\ \hline
$\mathbf{5g}\,\{5,4,9/2\} $ & $470.4016$ & $0.0004$ & $-0.2104 $&
$0.0457$ & $0.149206529721493\cdot 10^{-6}$
\\ \hline
$\mathbf{6p}\,\{6,1,1/2\} $ & $476.3844$ & $0.0038$ & $-0.2070 $&
$0.0460$ & $0.259039259820245\cdot 10^{-6}$
\\ \hline
$\mathbf{6p}\,\{6,1,3/2\} $ & $476.3866$ & $0.0016$ & $-0.2092 $&
$0.0460$ & $0.259039259820245\cdot 10^{-6}$
\\ \hline
$\mathbf{6s}\,\{6,0,1/2\} $ & $476.3845$ & $0.0037$ & $-0.2071 $&
$0.0459$ & $0.777119841553068\cdot 10^{-6}$
\\ \hline
$\mathbf{6d}\,\{6,2,3/2\} $ & $476.3866$ & $0.0016$ & $-0.2092 $&
$0.0460$ & $0.15542347340888\cdot 10^{-6}$
\\ \hline
$\mathbf{6d}\,\{6,2,5/2\} $ & $476.3873$ & $0.0009$ & $-0.2099 $&
$0.0460$ & $0.15542347340888\cdot 10^{-6}$
\\ \hline
$\mathbf{6f}\,\{6,3,5/2\} $ & $476.3873$ & $0.0009$ & $-0.2099 $&
$0.0460$ & $0.11101674147068\cdot 10^{-6}$
\\ \hline
$\mathbf{6f}\,\{6,3,7/2\} $ & $476.3877$ & $0.0005$ & $-0.2103 $&
$0.0460$ & $0.11101674147068\cdot 10^{-6}$
\\ \hline
$\mathbf{6g}\,\{6,4,7/2\} $ & $476.3877$ & $0.0005$ & $-0.2103 $&
$0.0460$ & $0.8634634356673\cdot 10^{-7}$
\\ \hline
$\mathbf{6g}\,\{6,4,9/2\} $ & $476.3879$ & $0.0003$ & $-0.2105 $&
$0.0460$ & $0.8634634356673\cdot 10^{-7}$
\\ \hline
$\mathbf{6h}\,\{6,5,9/2\} $ & $476.3879$ & $0.0003$ & $-0.2105 $&
$0.0460$ & $0.70647002692126\cdot 10^{-7}$
\\ \hline
$\mathbf{6h}\,\{6,5,11/2\} $ & $476.3881$ & $0.0001$ & $-0.2107 $&
$0.0459$ & $0.70647002692126\cdot 10^{-7}$
\\ \hline
$\mathbf{7p}\,\{7,1,1/2\} $ & $479.9953$ & $0.0024$ & $-0.2084 $&
$0.0461$ & $0.16312668406637\cdot 10^{-6}$
\\ \hline
$\mathbf{7p}\,\{7,1,3/2\} $ & $479.9967$ & $0.0010$ & $-0.2098 $&
$0.0461$ & $0.16312668406637\cdot 10^{-6}$
\\ \hline
$\mathbf{7s}\,\{7,0,1/2\} $ & $479.9954$ & $0.0023$ & $-0.2085 $&
$0.0461$ & $0.489380869959334\cdot 10^{-6}$
\\ \hline
\end{tabular}
\end{center}
\end{table}
 \mbox{}
\begin{table} []
\caption{Properties of the energy levels of the $H$-like atom,
$^{12}C$ ( the continuation of the Table 1).}
\begin{center}
\begin{tabular}{|c||c|c|c|c|c|}
\hline
 $ \, n,\,l,\,j $&$E_{n,l,j}^{exp.} $ & $\delta E_1$ &$\delta E_2$
 &$\delta E_3$ &$x_{nl}=\frac{\hbar c}{\varepsilon_1^2}\cdot f_{nl} $
 \\ \hline\hline
$\mathbf{7d}\,\{7,2,3/2\} $ & $479.9967$ & $0.0010$ & $-0.2098 $&
$0.0461$ & $0.9787597772952\cdot 10^{-7}$
\\ \hline
$\mathbf{7d}\,\{7,2,5/2\} $ & $479.9971$ & $0.0006$ & $-0.2102 $&
$0.0462$ & $0.9787597772952\cdot 10^{-7}$
\\ \hline
$\mathbf{7f}\,\{7,3,5/2\} $ & $479.9971$ & $0.0006$ & $-0.2102 $&
$0.0462$ & $0.69911402650592\cdot 10^{-7}$
\\ \hline
$\mathbf{7f}\,\{7,3,7/2\} $ & $479.9974$ & $0.0003$ & $-0.2105 $&
$0.0461$ & $0.69911402650592\cdot 10^{-7}$
\\ \hline
$\mathbf{7g}\,\{7,4,7/2\} $ & $479.9973$ & $0.0004$ & $-0.2104 $&
$0.0462$ & $0.5437553106815\cdot 10^{-7}$
\\ \hline
$\mathbf{7g}\,\{7,4,9/2\} $ & $479.9975$ & $0.0002$ & $-0.2106 $&
$0.0461$ & $0.5437553106815\cdot 10^{-7}$
\\ \hline
$\mathbf{7h}\,\{7,5,9/2\} $ & $479.9975$ & $0.0002$ & $-0.2106 $&
$0.0461$ & $0.44489068621169\cdot 10^{-7}$
\\ \hline
$\mathbf{7h}\,\{7,5,11/2\} $ & $479.9976$ & $0.0001$ & $-0.2107 $&
$0.0461$ & $0.44489068621169\cdot 10^{-7}$
\\ \hline
$\mathbf{7i}\,\{7,5,11/2\} $ & $479.9976$ & $0.0001$ & $-0.2107 $&
$0.0461$ & $0.37644595205933\cdot 10^{-7}$
\\ \hline
$\mathbf{7i}\,\{7,5,13/2\} $ & $479.9976$ & $0.0001$ & $-0.2107 $&
$0.0462$ & $0.37644595205933\cdot 10^{-7}$
\\ \hline
$\mathbf{8p}\,\{8,1,1/2\} $ & $482.3388$ & $0.0016$ & $-0.2092 $&
$0.0462$ & $0.109282103914287\cdot 10^{-6}$
\\ \hline
$\mathbf{8p}\,\{8,1,3/2\} $ & $482.3397$ & $0.0007$ & $-0.2101 $&
$0.0463$ & $0.109282103914287\cdot 10^{-6}$
\\ \hline
$\mathbf{8s}\,\{8,0,1/2\} $ & $482.3388$ & $0.0016$ & $-0.2092 $&
$0.0462$ & $0.327846678749163\cdot 10^{-6}$
\\ \hline
$\mathbf{8d}\,\{8,2,3/2\} $ & $482.3397$ & $0.0007$ & $-0.2101 $&
$0.0463$ & $0.65569247668352\cdot 10^{-7}$
\\ \hline
$\mathbf{8d}\,\{8,2,5/2\} $ & $482.3400$ & $0.0004$ & $-0.2104 $&
$0.0463$ & $0.65569247668352\cdot 10^{-7}$
\\ \hline
$\mathbf{8f}\,\{8,3,5/2\} $ & $482.3400$ & $0.0004$ & $-0.2104 $&
$0.0463$ & $0.46835172412023\cdot 10^{-7}$
\\ \hline
$\mathbf{8f}\,\{8,3,7/2\} $ & $482.3402$ & $0.0002$ & $-0.2106 $&
$0.0462$ & $0.46835172412023\cdot 10^{-7}$
\\ \hline
$\mathbf{8g}\,\{8,4,7/2\} $ & $482.3402$ & $0.0002$ & $-0.2106 $&
$0.0462$ & $0.36427354378636\cdot 10^{-7}$
\\ \hline
$\mathbf{8g}\,\{8,4,9/2\} $ & $482.3403$ & $0.0001$ & $-0.2107 $&
$0.0462$ & $0.36427354378636\cdot 10^{-7}$
\\ \hline
$\mathbf{8h}\,\{8,5,9/2\} $ & $482.3403$ & $0.0001$ & $-0.2107 $&
$0.0462$ & $0.29804198026032\cdot 10^{-7}$
\\ \hline
$\mathbf{8h}\,\{8,5,11/2\} $ & $482.3403$ & $0.0001$ & $-0.2107 $&
$0.0463$ & $0.29804198026032\cdot 10^{-7}$
\\ \hline
$\mathbf{8i}\,\{8,6,11/2\} $ & $482.3403$ & $0.0001$ & $-0.2107 $&
$0.0463$ & $0.25218936198996\cdot 10^{-7}$
\\ \hline
$\mathbf{8i}\,\{8,6,13/2\} $ & $482.3404$ & $0.0000$ & $-0.2108 $&
$0.0462$ & $0.25218936198996\cdot 10^{-7}$
\\ \hline
$\mathbf{8k}\,\{8,7,13/2\} $ & $482.3404$ & $0.0000$ & $-0.2108 $&
$0.0462$ & $0.21856410996047\cdot 10^{-7}$
\\ \hline
$\mathbf{8k}\,\{8,7,15/2\} $ & $482.3404$ & $0.0000$ & $-0.2108 $&
$0.0462$ & $0.21856410996047\cdot 10^{-7}$
\\ \hline
$\mathbf{9p}\,\{9,1,1/2\} $ & $483.9454$ & $0.0012$ & $-0.2096 $&
$0.0463$ & $0.76752301621132\cdot 10^{-7}$
\\ \hline
$\mathbf{9p}\,\{9,1,3/2\} $ & $483.9461$ & $0.0005$ & $-0.2103 $&
$0.0463$ & $0.76752301621132\cdot 10^{-7}$
\\ \hline
$\mathbf{9s}\,\{9,0,1/2\} $ & $483.9455$ & $0.0011$ & $-0.2097 $&
$0.0462$ & $0.230257085896465\cdot 10^{-6}$
\\ \hline
$\mathbf{9d}\,\{9,2,3/2\} $ & $483.9461$ & $0.0005$ & $-0.2103 $&
$0.0463$ & $0.46051373731367\cdot 10^{-7}$
\\ \hline
$\mathbf{9d}\,\{9,2,5/2\} $ & $483.9463$ & $0.0003$ & $-0.2105 $&
$0.0463$ & $0.46051373731367\cdot 10^{-7}$
\\ \hline
$\mathbf{9f}\,\{9,3,5/2\} $ & $483.9463$ & $0.0003$ & $-0.2105 $&
$0.0463$ & $0.32893836162821\cdot 10^{-7}$
\\ \hline
$\mathbf{9f}\,\{9,3,7/2\} $ & $483.9464$ & $0.0002$ & $-0.2106 $&
$0.0463$ & $0.32893836162821\cdot 10^{-7}$
\\ \hline
$\mathbf{9g}\,\{9,4,7/2\} $ & $483.9464$ & $0.0002$ & $-0.2106 $&
$0.0463$ & $0.2558409383546\cdot 10^{-7}$
\\ \hline
$\mathbf{9g}\,\{9,4,9/2\} $ & $483.9465$ & $0.0001$ & $-0.2107 $&
$0.0463$ & $0.2558409383546\cdot 10^{-7}$
\\ \hline
$\mathbf{9h}\,\{9,5,9/2\} $ & $483.9465$ & $0.0001$ & $-0.2107 $&
$0.0463$ & $0.20932439912118\cdot 10^{-7}$
\\ \hline
$\mathbf{9h}\,\{9,5,11/2\} $ & $483.9465$ & $0.0001$ & $-0.2107 $&
$0.0463$ & $0.20932439912118\cdot 10^{-7}$
\\ \hline
$\mathbf{9i}\,\{9,6,11/2\} $ & $483.9465$ & $0.0001$ & $-0.2107 $&
$0.0463$ & $0.17712064248878\cdot 10^{-7}$
\\ \hline
$\mathbf{9i}\,\{9,6,13/2\} $ & $483.9465$ & $0.0001$ & $-0.2107 $&
$0.0463$ & $0.17712064248878\cdot 10^{-7}$
\\ \hline
$\mathbf{9k}\,\{9,7,13/2\} $ & $483.9465$ & $0.0001$ & $-0.2107 $&
$0.0463$ & $0.15350455496686\cdot 10^{-7}$
\\ \hline
$\mathbf{9k}\,\{9,7,15/2\} $ & $483.9466$ & $0.0000$ & $-0.2108 $&
$0.0463$ & $0.15350455496686\cdot 10^{-7}$
\\ \hline
$\mathbf{9l}\,\{9,8,15/2\} $ & $483.9466$ & $0.0000$ & $-0.2108 $&
$0.0463$ & $0.13544519430618\cdot 10^{-7}$
\\ \hline
$\mathbf{9l}\,\{9,8,17/2\} $ & $483.9466$ & $0.0000$ & $-0.2108 $&
$0.0463$ & $0.13544519430618\cdot 10^{-7}$
\\ \hline
\end{tabular}
\end{center}
\end{table}

\begin{equation}
\label{50} \begin{array}{ll} E_{n,l}\Rightarrow E_{n,l}-E_{1,0}\,;
 \\
E_{n}^{Q.mech.}\Rightarrow E_{n}^{Q.mech}-E_{1}^{Q.mech}\,;\quad
E_{n,j}^{Dir.}\Rightarrow E_{n,j}^{Dir.}-E_{1,\frac12}^{Dir.}\, .
\end{array}
\end{equation}
In the 3-rd, 4-th and 5-th columns, there are given the values of
the deviations of the theoretical numbers for $E_{nl},
E_n^{Q.mech.}$, and $E_{nj}^{Dir.}$ in relation to the
corresponding experimental value $E_{n,l,j}^{exp.}$, more exactly,
\begin{equation}
\label{51}  \delta E_1\equiv E_{n,l}-E_{n,l,j}^{exp.};\quad \delta
E_2\equiv E_n^{Q.mech.}-E_{n,l,j}^{exp.};\quad \delta E_3\equiv
E_{n,j}^{Dir.}-E_{n,l,j}^{exp.};\, .
\end{equation}
Finally, in the last column of the Table the values of the root
$x_{nl}=\frac{\hbar c}{\varepsilon_1^2}\cdot f_{nl}$ of the
algebraic equation (33) for every $(n,l)$-level are given. The
specific values for $x_{nl}$ given therein provide the accuracy of
the solution of (33) $\sim 10^{-21}$.

As is seen from the Table I, the NOCE model yields the most
accurate description (average deviation from the experimental
values is $\overline{\delta E_1}\equiv\sum_i \delta
E_1^{(i)}/81=0.3120/81\simeq 0.00385$ eV). The Dirac theory is an
order of accuracy worse than the NOCE model ( the average
deviation $\overline{\delta E_3}\equiv\sum_i \delta
E_3^{(i)}/81=3.7010/81\simeq 0.04569$ eV). Of even less accuracy
is the description by the formula $E_n^{Q.mech.}$ ( here the
average deviation is $\overline{\delta E_2}\equiv\sum_i \delta
E_2^{(i)}/81=16.7521/81\simeq 0.20682$ eV).

Among the results to be considered here, the important one, in our
opinion, is the conclusion about the increase in the masses of the
interacting particles, the electron and $Z$-nucleus. For example,
while the electron mass in the Hydrogen atom is growing by
$0.000388$ eV only in relation to the mass of a free particle, the
increase in the electron mass in the case of the $H$-atom $^{12}C$
is rather significant. The indirect clue for the increase in the
masses of the interacting particles is present in the calculation
of the energy spectrum of the $H$-atom $^{12}C$ based the equation
(46) of the Dirac theory. Indeed, as is seen from the Tables 1 and
2, the deviation of the theoretical numbers from the experimental
values for $\Delta m_e c^2$ for all $(nj)$-states remains
approximately constant. Therefore, by shifting the ground state
$(1,\frac12)$, say, by the magnitude $\delta\overline{E}_3$ while
holding the positions of the excited states, one can, apparently,
essentially improve the description of the experimental situation.
In order to fit the experimental values by $E_{1,1/2}^{Dir.}$, it
is necessary to increase the electron mass in the 1-st term in the
right-hand side of equation (46) by $\Delta m_e c^2\approx \delta
E_3^{(1)}/\sqrt{1-\alpha^2 Z^2}=0.0466165$ eV, while keeping the
electron rest mass $m_e c^2$ in the second term of this equation
constant. It is clear that for the excited states this shift must
be insignificant in relation to the ground state. Exactly this
behavior of the electron mass, as dependent on the state of the
$H$-atom $^{12}C$ under consideration, is characteristic of the
NOCE model. In particular, in accordance with the NOCE model, the
ratio of the quantity  $\Delta m_e c^2$ for the ground $1s$-state
to that for the first excited $2p$-state equals
\begin{equation}
\label{52}  \frac{\Delta m_e c^2(1s)}{\Delta m_e
c^2(2p)}=\left[1-\frac{1}{1-2\,\gamma_{12}\cdot x_{10}}\right]\div
\left[1-\frac{1}{1-2\,\gamma_{12}\cdot x_{20}}\right] \approx
24.027;\,
\end{equation}
and the same ratio for the last $(9l)$-state is $\Delta m_e
c^2(1s)/\Delta m_e c^2(9l)\approx 12409.672$, i.e., for the highly
excited states the increase in the electron mass turns out to be
negligibly small as compared to $\Delta m_e c^2$ for the ground
state.

To conclude the examination of the energy spectrum of the levels
of $H$-like atoms, let us briefly dwell upon the question about
the orthogonality of the wave functions (26). In the NOCE model,
as can be easily seen from the relation (32), the dimensionless
parameter $a_o$ (Bohr radius) depends on the specific $(nl)$-state
under consideration. For this reason some functions
$\psi_{nlm}(\mathbf{r})$ may be, in general, nonorthogonal to one
another, i.e., the overlap integral
 $$ j_{nl,n^{'}l}\equiv \int
\psi_{nlm}(\mathbf{r})
\psi_{nlm}(\mathbf{r})\,d\mathbf{r}=\int_0^\infty
R_{nl}(r)\,R_{n^{'}l}(r)\,r^2dr \, $$ may be different from zero.
It is clear that this circumstance may cast some doubt on the
above obtained results of the NOCE model in the case that integral
$j_{nl,n^{'}l}$ would be markedly different from zero. In this
context, let us estimate the integral $j_{nl,n^{'}l}$, taking as
an example $1s$- and $2s$-states of the Hydrogen atom and the
$H$-atom $^{12}C$. In view of (28),  we write the necessary
overlap integral in the form $$j_{1s,2s}\equiv\int_0^\infty
R_{10}(r)\,R_{20}(r)\,r^2\,d\,r=\left(\frac{2}{a_o\,\tilde{a}_0}
\right)^{3/2}\left(\frac{1}{\tilde{a}_o}+\frac{1}{2a_o}\right)^{-3}
\left[1-\frac{3}{2a_o}\left(\frac{1}{\tilde{a}_o}+\frac{1}{2a_o}
\right)^{-1}\right],$$ where $$\tilde{a}_o=\frac{\hbar\,c}{\alpha
\varepsilon_1}\left[\left(1-\gamma_{12}\frac{2\hbar\,c}{\varepsilon_1^2}\cdot
f_{10}\right)^2+\xi\left(1-\gamma_{21}\frac{2\hbar\,c}{\varepsilon_1^2}\cdot
\xi^2\,f_{10} \right)^2\right],$$ $$a_o=\frac{\hbar\,c}{\alpha
\varepsilon_1}\left[\left(1-\gamma_{12}\frac{2\hbar\,c}{\varepsilon_1^2}\cdot
f_{20}\right)^2+\xi\left(1-\gamma_{21}\frac{2\hbar\,c}{\varepsilon_1^2}\cdot
\xi^2\,f_{20} \right)^2\right].$$ Substituting the numerical
values for the Hydrogen atom
$x_{10}\equiv\frac{\hbar\,c}{\varepsilon_1^2}f_{10}\simeq
0.776342\cdot10^{-6},\,x_{20}\equiv\frac{\hbar\,c}{\varepsilon_1^2}f_{20}\simeq
0.970373\cdot 10^{-7}$ and for the $H$-atom $^{12}C$,
$x_{10}\approx 0.168002\cdot10^{-3},\,x_{20}\approx 0.209844\cdot
10^{-4}$ (see the tables 1,2) yields for the integral
$j_{1s,2s}\approx 3.709\cdot 10^{-10}$ for the case of the
Hydrogen atom and $j_{1s,2s}\approx 0.000210$ for the case of the
$H$-atom $^{12}C$. These estimates strongly evidence that such a
negligibly small value of the overlapping between the $1s$- and
$2s$-states may hardly considerably affect the positions of these
energy levels within the framework of the NOCE model.

\section{Conclusion} \label{sect4}

From the analysis performed in this work, we conclude that

1) When then particles of a microsystem interact with each other
by means of a force, the postulate about the noncommutativity of
the coordinate and impulse operators for a single particle
necessarily effects in the noncommutativity of the coordinate and
impulse operators of different particles. The generalization of
the basic principle of quantum mechanics presented in this work
leads us to the emergence of completely new physical behavior. One
of the examples for the above mentioned is the dependence of the
particle mass on the force of its interaction with other
particles. The theory under consideration, as is shown in this
work, establishes the limit for the matrix element of the force
\begin{equation}
\label{53}  f_o < \frac{1}{2\gamma_o}\cdot
\frac{(m_o\,c^2)^2}{\hbar c} \, ,
\end{equation}
beyond which the notion of "a particle" loses its sense. In this
connection, an interesting analogy  with the special relativity
comes to mind, where the particle energy  $T\equiv m_o\,v^2/2$ may
not exceed the value $m_o\,c^2/2$ and the value $m_o^2\,c^4$
entering (92) is connected with the invariance of the 4-vector of
the energy-impulse with respect to the Lorentz transformation. It
is advisable to remember that the Dirac theory yields the known
limitation on the magnitude of the force of interaction between
charged particles $(Z_{crit.}\sim 137)$ as well. Within our
approach, the Coulomb interaction represents an ordinary example
of the interaction force.

\centerline{}

\end{document}